\documentclass[referee]{aa}

\usepackage{graphicx}
\usepackage{natbib}
\usepackage{txfonts}
\input amssym.tex

\begin{document}

\title
  {Testing
  the universal stellar IMF on the metallicity distribution in the
  bulges of the Milky Way and M31}

\author
{S.K. Ballero\inst{1,2}
\thanks {E-mail: ballero@oats.inaf.it}
\and  P. Kroupa\inst{3} 
\and  F. Matteucci\inst{1,2}}

\institute{Dipartimento di Astronomia, Universit\`a di Trieste,
    Via G.B. Tiepolo 11, I-34121 Trieste, Italy
\and INAF-Osservatorio Astronomico di Trieste, Via G.B. Tiepolo 11,
I-34121 Trieste, Italy
\and Argeland Institute for Astronomy, University of Bonn, Auf dem
H\"ugel 71, 53121 Bonn, Germany}

\date{Received xxxx, Accepted xxxx}

\abstract{}
{We test whether the universal initial mass
function (UIMF) or the integrated galaxial IMF (IGIMF) can be
employed to explain the metallicity distribution (MD) of giants in the
Galactic bulge.}
{We make use of a single-zone chemical evolution model
developed for the Milky Way bulge in the context of an inside-out
model for the formation of the Galaxy.
We checked whether it is possible to constrain the yields above $80
M_{\sun}$ by forcing the UIMF and required that the resulting MD
matches the observed ones.
We also extended the analysis to the bulge of M31 to investigate a
possible variation of the IMF among galactic bulges.
Several parameters that have an impact on stellar
evolution (star-formation efficiency, gas infall timescale)
are varied.}
{We show that it is not possible to satisfactorily reproduce the
observed metallicity distribution in the two galactic bulges unless
assuming a flatter IMF ($x \leq 1.1$) than the universal one.}  
{We conlude that
it is necessary to assume a variation in the IMF among the
various environments.}  


\titlerunning{The Universal IMF in the bulges of Milky Way and M31}

\maketitle

\authorrunning{S.K. Ballero et al.}

\section{Introduction}
The question of which is the most suitable initial mass distribution
for bulges of galaxies is not addressed very often. In fact,
bulge evolution models (e.g. Samland et al. 1997; Ferreras et al.
2003; Immeli et al. 2004; Costa et al. 2005) usually
assume \emph{a priori} that the zero age main sequence masses 
of stars are distributed following a power-law
distribution: 
\begin{equation}
\phi(m) \propto m^{-(1+x)}
\end{equation}
with a Salpeter (1955) index ($x=1.35$).
Basic physical arguments support the idea of an initial mass
function (IMF) varying among different environments (see e.g.
Padoan et al. 1997; Larson 1998; or Nakamura \& Umemura
2001; Schaerer 2002; Bromm \& Larson 2004 for Population III~stars).

On the other hand, so far there has not been a convincing observational
evidence of such a variation in the stellar IMF based on direct
stellar counts (see e.g. Chabrier 2003 for an extensive review).
Massey (1998) find that the IMF is well represented by a Salpeter
slope over an order of magnitude in metallicity, in the clusters and
associations of the Milky Way and Magellanic Clouds, as well as in OB
associations, while the slope appears to be much steeper
in the field ($x \sim -3$). 
The invariance of the stellar IMF was confirmed by subsequent works. 
Kroupa (2001) summarized the available constraints by means of the  
multi-part power-law shape
\begin{equation}
  \phi(M)\propto M^{-(1+x_i)}
\end{equation}
where
\begin{equation}
  \begin{array}{l}
x_1 = 0.3 \mbox{ for } 0.08 \leq M/M_{\odot} \leq 0.5\\
x_2 = 1.3 \mbox{ for } 0.5\phantom{0} \leq M/M_{\odot}
  \end{array}
\label{eq1}
\end{equation}
which we call the Universal IMF (UIMF).\footnote{This 
IMF is the form corrected for
unresolved binaries.}

However, the IMF integrated over galaxies, which controls the
distribution of stellar remnants, number of supernovae (SNe), and the
chemical enrichment of a galaxy, is generally different from the
stellar IMF and is given by the integral of the latter over the
embedded star-cluster mass function, which varies from galaxy to
galaxy.
Weidner \& Kroupa (2005) find such integrated galaxial IMF (IGIMF) to
be steeper than the UIMF for a range of plausible scenarios, and they 
suggest a ``maximum scenario'', based on the Scalo (1986) star-count
analysis of the local Galactic field, with an IGIMF which has the
following indexes:
\begin{equation}
  \begin{array}{l}
x_1 = 0.3 \mbox{ for } 0.08 \leq M/M_{\odot} \leq 0.5\\
x_2 = 1.3 \mbox{ for } 0.5\phantom{0} \leq M/M_{\odot} \leq 1\\
x_3 = 1.7 \mbox{ for } 1\phantom{.00} \leq M/M_{\odot}
  \end{array}
\label{eq2}
\end{equation}
In the following, we will refer to this IMF as to the IGIMF.

Conversely, Piotto \& Zoccali (1999) and Paresce \& De Marchi (2000)
measured the present-day mass function in Galactic globular
  clusters below $\sim 0.7 M_{\odot}$. 
They found evidence of variation in the MF slope among different
environments, with a tendency toward flatter slopes in globular clusters
compared to the Galactic field IMF; Paresce \& De Marchi (2000)
also state that, in the considered mass range, the observed mass
function represents the true stellar IMF for these environments.
Zoccali et al. (2000) derived the IMF below $1M_{\odot}$ in the
Galactic bulge and concluded that it is shallower than the Salpeter
slope; it also shows similarity with the IMF of globular clusters.
However, the somewhat smaller $x_1$ is probably the result of the
evaporation of low-mass stars from the cluster (Baumgardt \& Makino
2003), while the Bulge result is still consistent with the UIMF,
within the uncertainties.

Concerning the range of masses over which star formation is possible,
although stellar instabilities that potentially lead to disruption
already occur above $60-120 M_{\odot}$ (Schwarzschild \& H\"arm 1959),
stars of $\sim 140-155 M_{\odot}$ were observed in the core of the R136
cluster (Massey \& Hunter, 1998) and in the Arches cluster (Figer,
2005).
Although Massey (2003) sustains that this upper limit may indeed be
statistical rather than physical (i.e. there may not be regions that
are rich enough to allow the detection of such stars), Oey \& Clarke
(2005) studied the content of massive stars in 9 clusters and OB
associations in the Milky Way, LMC, and SMC and find that the
expectation value for the maximum stellar mass lies, with high
significance, in the range $120-200M_{\odot}$. 
This agrees with the conclusion of Weidner \& Kroupa (2004) that a
fundamental maximum limit for stellar masses can be constrained at
about $150 M_{\odot}$, unless the true stellar IMF has $x > 1.8$. 

Attempts to constrain the IMF in the bulge of our galaxy based on
observations of chemical abundances were carried out by Matteucci \& 
Brocato (1990) and Matteucci et al. (1999), who fixed the index by 
the requirement of reproducing the observed metallicity
distributions (MDs) of Rich (1988) and McWilliam \& Rich (1994),
respectively.
They both concluded that the bulge IMF must be flatter than the
Salpeter one, in general, and lie in the range $x= 1.1-1.35$, thus
favoring the production of massive stars with respect to the solar
vicinity.
An even flatter IMF index was chosen by Ballero et al. (2006b), who
showed that it is necessary to assume $x_2=0.95-1.1$ 
(where $x_2$ is the one defined in Eq. \ref{eq1}) to fit the MDs of
Zoccali et al. (2003) and Fulbright et al. (2006) for the Galactic
bulge and of Sarajedini \& Jablonka (2005) for the bulge of M31.
Even shallower IMFs ($x=0.33$) are compatible with these observed
distributions, but give rise to a certain amount of oxygen
overproduction.

Our present aim is to test the effect of adopting both the UIMF and
the IGIMF on the predicted bulge MD and to compare the results with
the MDs employed in Ballero et al. (2006b).
These IMFs will be extended to a much higher upper mass limit than in
previous models, so we will try to find the combination of metal
yields and evolutionary parameters that best fit the observations
with these two IMFs.

In \S 2 we present the adopted chemical evolution models, in \S 3 we
discuss the outcome of these models and the results of their
comparisons with the observed MDs, and in \S 4 we draw some conclusions.

\section{The chemical evolution models}

We briefly summarize the chemical evolution model described in Ballero et
al. (2006b), which we~use.
The star formation rate (SFR) is parametrized
as $\psi (r,t)=\nu G^k(r,t)$, where $\nu$ is the star-formation
efficiency, i.e. the inverse of the timescale of star formation, $k=1$
to recover the star formation law of spheroids (but it can be
shown that remarkable differences do not arise with $k=1.5$), and
$G(r,t)$ is the gas surface mass density $\sigma_{gas}$ normalized to
the present time value. 
The bulge forms by accreting gas from the halo at a rate $\dot{G}(t)
\propto e^{-t/\tau}$, where $\tau$ is the collapse timescale. 
The metallicity $Z_{acc}$ of the accreted gas is very low, 
and it can be shown that the results do not change significantly if we
consider $Z_{acc} \simeq 0$.

The Type Ia SN rate is computed according to Matteucci \&
Recchi (2001) following the single degenerate scenario of Nomoto et
al. (1984).
Stellar lifetimes (Kodama 1997) are taken into account in
detail; nucleosynthesis prescriptions are taken from Fran\c cois et 
al. (2004), who constrained the stellar yields in order to reproduce
the chemical properties of the solar neighbourhood via the two-infall
model of Chiappini et al. (2003).
The gas is supposed to be well-mixed and homogeneous at any time.
The binding energy of the Galactic bulge (contributed by the bulge
itself and the dark matter halo), as well as the thermal energy
injected by SNe, is calculated as in Matteucci (1994); when the
thermal energy equals the binding energy, the star formation is highly
suppressed, even though the gas remains bound to the
bulge itself after this occurs.
This event does not have a great impact on the predicted MD, since in
any case it occurs when most of the gas 
has already been processed into stars. 
However, it helps avoid
overestimating the high-metallicity tail of the MD.
Finally, the adopted IMF has the general shape of a multi-part
power law:
\begin{equation}
  \phi(M)\propto M^{-(1+x_i)}
\end{equation}
where the subscript refers to different mass ranges.

The reference model of Ballero et al. (2006b), which best fits the MD
and the [$\alpha$/Fe] \emph{vs.} [Fe/H] ratios in the Bulge,
has $\nu=20$ Gyr$^{-1}$, $\tau=0.1$ Gyr, and
a two-slope IMF, namely $x_1 = 0.33$ for $0.08 \leq M/M_{\odot} \leq 1$
(in agreement with the photometric measurements of Zoccali et al.,
2000) and $x_2 = 0.95$ for $1 \leq M/M_{\odot} \leq 80$.   

\begin{figure}
\centering
\includegraphics[width=.4\textwidth]{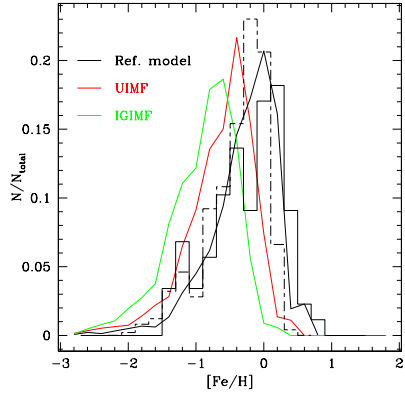}
\caption{[Fe/H] distributions calculated with the adoption of
  different IMFs. The solid line represents the fiducial model of
  Ballero et al. (2006b). The data are compared with the observed
  distributions of Zoccali et al. (2003, dashed histogram) and
  Fulbright et al. (2006, solid histogram). This figure excludes the
  IGIMF as a plausible one for the Galactic bulge.}
\label{fig1}
\end{figure}

\begin{figure}
\centering
\includegraphics[width=.4\textwidth]{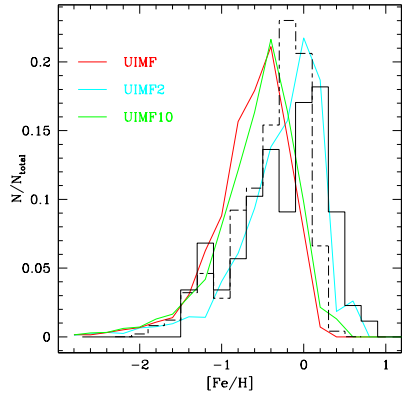}
\caption{[Fe/H] distributions obtained with the UIMF and the adoption
  of different yields for massive stars (see text for details),
  compared to the same observed distributions of the previous figure.
  A huge variation in the Fe yields above $80 M_{\odot}$ leads to a
  negligible shift in the MD.}
\label{fig2}
\end{figure}

\section{Model results}

In Ballero et al. (2006b) we showed that an IMF with the
Salpeter (1955) index ($x=1.35$) for $1 \leq M/M_{\odot} \leq 80$ or
with the Scalo (1986) index ($x=1.7$) for $2 \leq M/M_{\odot} \leq 80$
could not fit the observed MDs in any way, causing the resulting
distribution to be shifted towards low metallicities. 
Now we test whether an extension of these IMFs to a wider range
of masses can provide an Fe enrichment sufficient to shift the
calculated MD to the suitable position.

Figure \ref{fig1} compares the MD obtained with
the reference model and with the adoption of the UIMF and IGIMF,
as described by Eqs. \ref{eq1} and \ref{eq2}, and with all the other
parameters kept constant with respect to the reference model. 
The yields of Fran\c cois et al. (2004) have been extrapolated above
$80M_{\odot}$ by freezing the yields. 
It is quite evident that the UIMF
reproduces neither of the observed MDs, being shifted to metallicities
that are too low. 
Even worse results are obtained with the IGIMF, which has $<$[Fe/H]$>$
$\simeq -0.9$, i.e $\simeq 0.5-0.7$ dex lower than the observed 
ones. 

We then tried to force the UIMF by changing the yields
above $80M_{\odot}$, which have not been constrained so far. 
We found out (see Fig. \ref{fig2}) that, even if we adopt stellar
yields as large as ten times the extrapolated ones (Model UIMF10),
which is rather unrealistic, the predicted MD is only negligibly
affected.  
Only if we double all the Fe yields from massive stars ($M >
10M_{\odot}$, Model UIMF2) does the calculated MD become consistent
with the observations, but this would require also doubling the
yields of other elements in order to preserve the agreement with
e.g. the [$\alpha$/Fe] vs. [Fe/H] plots (see Ballero et al., 2006b),
and this is theoretically implausible.
Also, the adoption of these Fe yields for the solar neighbourhood would
destroy the agreement of the two-infall model with the observed solar
vicinity MD (e.g. Chiappini et al. 2003; Hou et al. 2000).

The main contributors to the Fe enrichment in the bulge are Type II
SNe, which have massive progenitors, since the timescales of
enrichment are so short. 
As a consequence of that, Ballero et al. (2006b) predict
overabundance of $\alpha$-elements relative to Fe for a wide range of
[Fe/H].
It should be possible to obtain a larger enrichment from
Type Ia SNe, which originate from low- and intermediate- mass stars,
by adjusting other parameters that have an effect on chemical
evolution, such as the star-formation efficiency $\nu$ or the infall
timescale $\tau$. 
Therefore, we also investigated two more models: Model
UIMF-$\nu$A, where we set $\nu=5$ Gyr$^{-1}$, and Model UIMF-$\tau$,
where $\tau=0.25$ Gyr. The star formation and gas consumption in
  this case should be slower, giving Type Ia SNe more time to 
enrich the interstellar medium with Fe. 
However, it should be possible to enhance the Fe
production by increasing the contribution of Type II SNe with a
faster enrichment, i.e. increasing the star-formation efficiency;
therefore, we also considered Model UIMF-$\nu$B, where $\nu=50$
Gyr$^{-1}$.

\begin{figure}
\centering
\includegraphics[width=.4\textwidth]{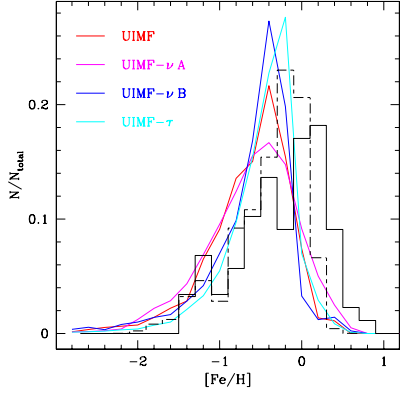}
\caption{Comparison of the observed MDs with those calculated with the
  UIMF and with models that adopt different values for the
  star-formation efficiency (UIMF-$\nu$A and B) and the gas infall 
  timescale (UIMF-$\tau$). Namely, higher values were chosen for
  $\tau$ and both higher and lower values for $\nu$.}
\label{fig3}
\end{figure}

We see in Fig. \ref{fig3}, however, that the attemp to shift the
position of the  MD by means of a change in these parameters is not
successful, as already shown in Ballero et al. (2006b) since they
mainly act on the broadness of the distribution.

In the case of a very high star-formation efficiency, the effect
can be explained if we consider that gas consumption occurs very 
rapidly and that stars are no longer formed after a very short time. 
  On the other hand, since the enrichment is very fast, there is a
  lack of metal-poor stars.
  The opposite occurs in the case of a lower star-formation
  efficiency.
In the case of different timescales of infall, the peak is actually
shifted, as can be seen from the figure, but this shift it is not
useful since the correct shape of the MD is not preserved. 
  This is because, if the gas accretes more slowly, the number of stars
  produced at low metallicities is lower, so the
  calculated MD gets sharper.

Finally, Fig. 4 shows the MDs resulting from our reference model and
the model with the UIMF compared to the MD of M31 as measured by
Sarajedini \& Jablonka (2005) translating the observed color-magnitude
diagram at $\sim 1.6$ kpc (G170 bulge field) from the center into a
MD function by means of red giant branches with various metallicities.
This MD, though still consistent with the ones of the Galactic bulge
and therefore indicating a similar enrichment history of the two
bulges, is slightly more metal-poor on average and is compatible
with both the reference model and the model with the UIMF; 
however, we also
plotted the MD calculated with $x_2=1.1$, like in Matteucci \&
Brocato (1990, model MB90), which gives the best fit. 
In any case, we can safely conclude that an IMF with $x \sim
1$ is more suitable for galactic bulges than the UIMF.

\begin{figure}
\centering
\includegraphics[width=.4\textwidth]{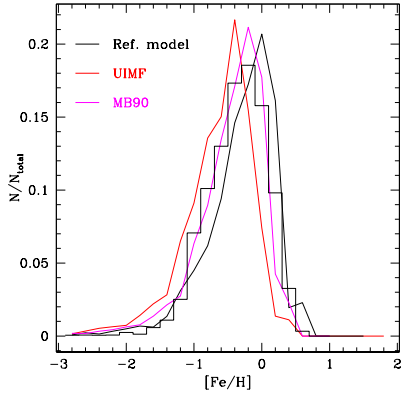}
\caption{[Fe/H] distribution function for the G170 field of the M31
  bulge field G170 (histogram) measured by Sarajedini \& Jablonka
  (2005) compared with the results of our reference model 
  and of the UIMF model. 
  We also show the results of model MB90, which is
  intermediate between the two and provides the best fit to the
  observed MD.}
\label{fig4}
\end{figure}

Other possibilities that could affect the
chemical evolution have not been investigated.
The lower mass cutoff of the IMF is constrained by measurements in
the bulge field and cluster giant stars (Kuijken \& Rich 2002;
Zoccali et al. 2003; Rich \& Origlia 2005) that indicate that the
bulk of them is roughly 10 Gyr old; therefore, the value of
the lowest mass cutoff cannot be higher than $\sim1 M_{\odot}$.
No difference is expected to arise if we increase $M_{inf}$ to that
value, since stars below $1 M_{\odot}$ have not yet contributed to the
bulge enrichment.
Furthermore, it has already been shown by Ballero et al. (2006a) that the
adoption of a mass cutoff of $10M_{\odot}$ for the so-called
Population III stars up to a metallicity suitable for the formation of
these stars (i.e. $Z \simeq 10^{-8}-10^{-4}$; Bromm \& Larson 2004)
has almost no effect on the predicted MD, and even less of one in the
bulge since such metallicities are reached in a very short time.

Lowering the SFR with the UIMF/IGIMF models in the bulge and accreting
supersolar-metallicity populations, such as are evident in the ancient
super-metal-rich open cluster NGC6791 (Salaris et al. 2004), or
accreting supersolar gas may be able to match the observed MD in the
bulges of the Milky Way and M31.
This scenario would avoid changing the IMF, but would also imply
substantial accretion events in the build-up of bulges. 
Such accretion episodes are likely to modify the
[$\alpha$/Fe] \emph{vs.} [Fe/H] plots at variance with observations. 
Due to the stochastic nature of mergers, each of them introduces a
wide spread in the plots, which is not observed.
Continuous outflows have the effect of lowering the effective yields
(see Tosi et al., 1998) and thus cannot be invoked to reproduce the
observed MD with the adoption of the UIMF.

\section{Conclusions}

We have tested the possibility of the UIMF of Kroupa (2001) or the
``maximal'' IGIMF of Weidner \& Kroupa (2005) holding in the bulge of
our galaxy and of M31.   
To this purpose, we included those IMFs in the chemical evolution
model of Ballero et al. (2006b), which reproduces the properties
of the Galactic bulge well. 
The upper mass limit was extended to $150 M_{\odot}$ in agreement with
late findings (Weidner \& Kroupa 2004; Oey \& Clarke 2005; Figer
2005; Koen 2006), and the
stellar yields of Fran\c cois et al. (2004) were extrapolated up to
that mass.
An attempt to constrain the yields above $80 M_{\odot}$ by assuming
\emph{a priori} the validity of the UIMF was also made, and
other parameters such as the star-formation efficiency or infall
timescale were varied in order to achieve a better fit.

\begin{table}
\centering
\begin{tabular}{lrrr}
\hline
Observed distribution & $M$ & $\mu$ & $\sigma$\\
\hline
Zoccali et al. (2003, MW)         & $-0.2$ & $-0.397$ & $0.444$\\
Fulbright et al. (2006, MW)       & $+0.2$ & $-0.248$ & $0.523$\\
Sarajedini \& Jablonka (2005, M31) & $-0.2$ & $-0.369$ & $0.438$\\
\hline
Models\\
\hline		
Ref. Model  & $+0.0$ & $-0.297$ & $0.502$\\
UIMF        & $-0.4$ & $-0.632$ & $0.501$\\
IGIMF       & $-0.6$ & $-0.896$ & $0.494$\\
UIMF2       & $-0.4$ & $-0.580$ & $0.493$\\
UIMF10      & $+0.0$ & $-0.272$ & $0.514$\\
MB90        & $-0.2$ & $-0.447$ & $0.493$\\
UIMF-$\nu$A & $-0.4$ & $-0.481$	& $0.587$\\
UIMF-$\nu$B & $-0.4$ & $-0.599$	& $0.503$\\
UIMF-$\tau$ & $-0.2$ & $-0.550$ & $0.417$\\
\hline
\end{tabular}
\caption{Statistical properties of the measured (upper part) and
  calculated (lower part) metallicity distributions of galactic
  bulges.}
\label{tab1}
\end{table}

Table \ref{tab1} summarizes the statistical properties of the observed
and calculated distributions. The first column shows the reference or
the model name; the second column reports the position of the
  peak on the [Fe/H] axis, i.e. the mode $M$; and in the third and
  fourth columns the average $\mu$ and the standard deviation
  $\sigma$ of the considered distribution are shown, respectively. 
Together with the reference, it is indicated whether it applies to the
Milky Way (MW) or M31 bulge.
It was found that it is not possible to satisfactorily reproduce
satisfactory way the observed MDs of the Galactic bulge and of M31
with the UIMF, which has a Salpeter (1955) index above
$0.5M_{\odot}$, because the predicted MD is too metal-poor. 
The adoption of the IGIMF, which has a Scalo (1986) index above $1
M_{\odot}$, worsens the agreement further.
This highlights the fact that the main Fe contributors in galactic
bulges are Type II SNe, since the timescales of enrichment are so
short.
Changing the nucleosynthesis prescriptions does not have remarkable
effects, unless very unrealistic assumptions about the stellar yields
of all massive stars are made. 
Even dramatic changes of yields above $80 M_{\odot}$ do not
practically affect the calculated distribution. 
We thus do not exclude the possibility of a higher mass-cutoff, but 
show that it is impossible to put constraints on it based on chemical
abundances, since the weight in the IMF of stars in the mass range $80
< M/M_{\odot} < 150$ is negligible.

Changes in $\nu$ and $\tau$ do not lead to any improvement because
they mainly have an effect on the breadth of the distribution and
not on the position, which is governed by the adopted IMF.
This clearly indicates that, if the bulges of the Galaxy and M31
formed inside-out through the accretion of very metal-poor gas, a
variation in the stellar IMF is necessary among different environments
and that an IMF index around $x\sim1$, flatter than that of the UIMF,
is preferable for galactic bulges. 
Theoretically speaking, it can be explained if we note that the star
formation in bulges proceeds like in a burst (see Elmegreen 1999); 
there are suggestions in the literature (e.g. Baugh et al. 2005;
Nagashima et al. 2005; Okamoto et al. 2005) about a
top-heavy IMF in starbursts. 
Figer (2005) also finds a flat IMF in the Arches cluster near the
Galactic center (however, see also Kim et al. 2006).

\subsection*{Acknowledgements}

SKB wishes to thank Ata Sarajedini for providing material in a very 
timely manner.

\end{document}